\documentclass[twocolumn,showpacs,preprintnumbers,amsmath,amssymb]{revtex4}

\usepackage{bm}
\usepackage[colorlinks=true,linkcolor=blue,citecolor=blue,urlcolor=blue]{hyperref}
\usepackage{color}

\usepackage{graphicx}
\usepackage{epstopdf}

\begin{document}


\title{Lorentz TEM imaging of stripe structures embedded in a soft magnetic matrix}

\author{M. A. Basith}
\altaffiliation[Present address: ]{Department of Physics, Bangladesh University of Engineering and Technology, Dhaka-1000, Bangladesh.}
\affiliation{School of Physics \& Astronomy, University of Glasgow, G12 8QQ, United Kingdom.
}

\author{S. McVitie}
\affiliation{School of Physics \& Astronomy, University of Glasgow, G12 8QQ, United Kingdom.
}

\author{T. Strache}
\affiliation{Institute of Ion Beam Physics and Materials Research, Helmholtz-Zentrum Dresden - Rossendorf, 01314 Dresden, Germany}

\author{M. Fritzsche}
\affiliation{Institute of Ion Beam Physics and Materials Research, Helmholtz-Zentrum Dresden - Rossendorf, 01314 Dresden, Germany}

\author{A. Muecklich}
\affiliation{Institute of Ion Beam Physics and Materials Research, Helmholtz-Zentrum Dresden - Rossendorf, 01314 Dresden, Germany}

\author{J. Fassbender}
\affiliation{Institute of Ion Beam Physics and Materials Research, Helmholtz-Zentrum Dresden - Rossendorf, 01314 Dresden, Germany}

\author{J. McCord}
\altaffiliation[Present address: ]{Institute for Materials Science, University of Kiel, 24143 Kiel, Germany.}
\affiliation{Institute of Ion Beam Physics and Materials Research, Helmholtz-Zentrum Dresden - Rossendorf, 01314 Dresden, Germany}


\date{\today}

\begin{abstract}
N\'eel walls in soft magnetic NiFe/NiFeGa hybrid stripe structures surrounded by a NiFe film are investigated by high resolution Lorentz transmission electron microscopic imaging. An anti-parallel orientation of magnetization in 1000 $nm$ wide neighboring unirradiated-irradiated stripes is observed by forming high angle domain walls during magnetization reversal. Upon downscaling the stripe structure size from $1000 \ nm$ to $200 \ nm$ a transition from a discrete domain pattern to an effective magnetic medium is observed for external magnetic field reversal. This transition is associated with vanishing ability of hosting high angle domain walls between adjacent stripes. The investigation also demonstrated the potentiality of Lorentz microscopy to image periodic stripe structures well under micron length-scale.  
   
\end{abstract}

\maketitle
\section{Introduction} \label{I}
Magnetic patterning on a micro- and nanometer scale is of intense research interest for potential applications in sensors \cite{ref81} and bit patterned media for high density data storage applications \cite{ref82}. By lateral magnetic patterning of ferromagnetic thin films using ion irradiation, the magnetic properties such as saturation magnetization \cite{ref1,ref51,ref2,ref3, ref61} and anisotropy \cite{ref4,ref5,ref6, ref62} can be changed on a local scale. If the patterning size is chosen below certain intrinsic magnetic length scales, the patterned medium may show properties not observed in neither the original film nor in an equivalent film with changed magnetic properties. Ion irradiation has been used to pattern periodic ferromagnetic (FM) stripe array structures where FM stripes of different saturation magnetization are present \cite{ref2, ref64}. Such a periodic stripe like pattern of the ion-beam-modified films \cite{ref64,ref63} has proven already the strong influence of the lengthscale of the lateral patterning on the micromagnetic behavior of the hybrid material \cite{ref2,ref3}. Compared to conventional techniques like reactive ion etching, ion-beam assisted patterning provides additional advantages like the possibility to locally modify magnetic properties \cite{ref71,ref83} without multistep deposition processes using different materials \cite{ref64}.  \\

In this paper we look at irradiated FM stripes which have lateral dimensions on the micron and sub-micron length scale. This is significant as one of the fundamental length scales which can be used to initially explore the modified magnetic structures is that of the domain wall width. In thin films the width associated with the tails of a N\'eel wall extend over several $\mu m$ in soft magnetic $Ni_{81}Fe_{19}$ thin films \cite{ref7}. During external field reversal along the stripe axis, densely packed nearly $180^{\circ}$ inter-stripe domain walls are created for stripe width equal and above $1 \ \mu m$. These inter-stripe domain walls are even able to mediate an exchange spring behavior on a lateral basis \cite{ref2}. The N\'eel wall width does not depend only on the material properties but also on the geometry and the lateral dimensions \cite{ref9} of the microstructure where it is located. Here we report on magnetic hybrid structures consisting of arrays of periodic stripes which were produced by magnetic patterning, where the patterning was realized by local irradiation of the film by $Ga^+$ focused ion beam (FIB) irradiation \cite{ref6}.  The aim of the magnetic patterning was to produce localised regions with reduced magnetization. The influence of the downscaling of the patterning size on the domain wall formation and the magnetization reversal process in saturation magnetization modulated stripe structures are studied by structural and quantitative magnetic imaging in the transmission electron microscope (TEM). Furthermore an explanation of the magnetic behavior of the stripe patterns is investigated using micromagnetic modelling.

\section{Experimental details} \label{II}
\begin{figure}[hh]
\centering
\includegraphics[width=8cm]{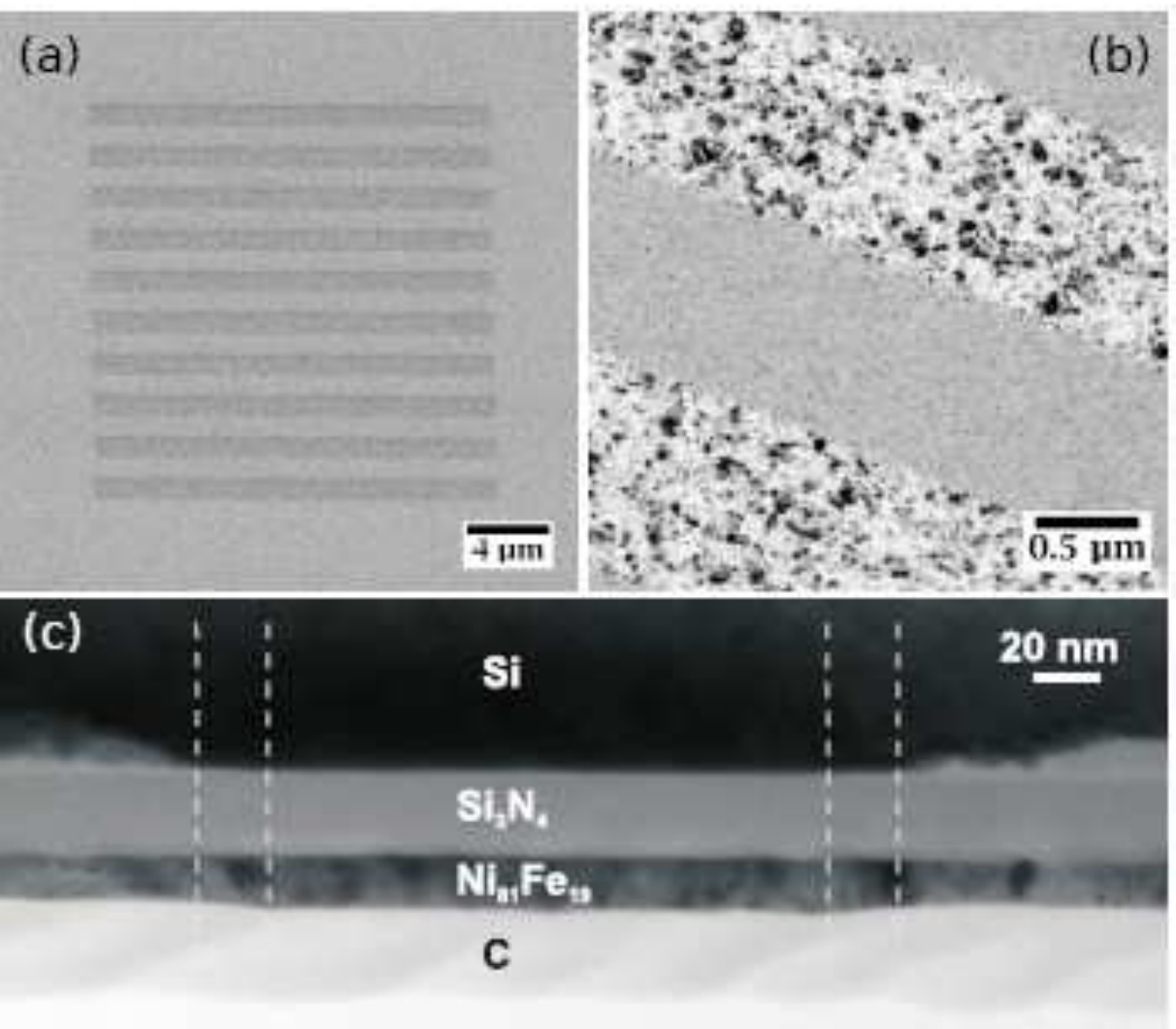}
\caption{Scanning electron microscopy image (a) of FIB written stripe array and plan-view bright field TEM image (b). The stripe width is in both cases $1 \ \mu m$. Cross-sectional TEM image (c) of an unirradiated area between two FIB irradiated stripes whose edges are indicated by the white dashed lines (stripe width $200 \ nm$).} \label{fig1}
\end{figure}

For the experiments polycrystalline $Ni_{81}Fe_{19}$ (atomic percent) films of nominal $20 \ nm$ 
thickness have been deposited by magnetron sputtering on top of $Si_{3}N_{4}$ window membranes \cite{ref10}, suitable for TEM investigations. During the deposition of the magnetic material an external magnetic field was applied to produce a magnetization induced magnetic anisotropy. Magnetic hybrid structures consisting of arrays of periodic stripes structures, as shown in  \textit{figure \ref{fig1}}, were produced by magnetic patterning. The aim of the magnetic patterning was the local reduction of the saturation magnetization of the material. The long axis of the stripes was aligned parallel to the anisotropy axis of the $Ni_{81}Fe_{19}$ film. The patterning was realized by $Ga^+$ focused ion beam (FIB) irradiation based lithography. A typical pattern is shown in a scanning electron microscope image (SEM) in \textit{figure \ref{fig1} (a)} where the film exhibits irradiated rectangular areas of dimensions of 1 $\mu m$ by 20 $\mu m$. The focused ion beam irradiation parameters were: energy of $30 \ keV$ $Ga^+$, current of $10 \ pA$, ion dose of $1000 \ \mu C/cm^2$ ($6.24$x$10^{15} \ ions/cm^2$), step size of $6 \ nm$, dwell time of $0.036 \ ms$ resulting in a beam speed of $0.167 \ mm/s$. Each stripe was irradiated once without any repetition. The long axis of the stripes was aligned parallel to the anisotropy axis of the $Ni_{81}Fe_{19}$ film. 

\begin{figure}[!hh]
\centering
\includegraphics[width=8cm]{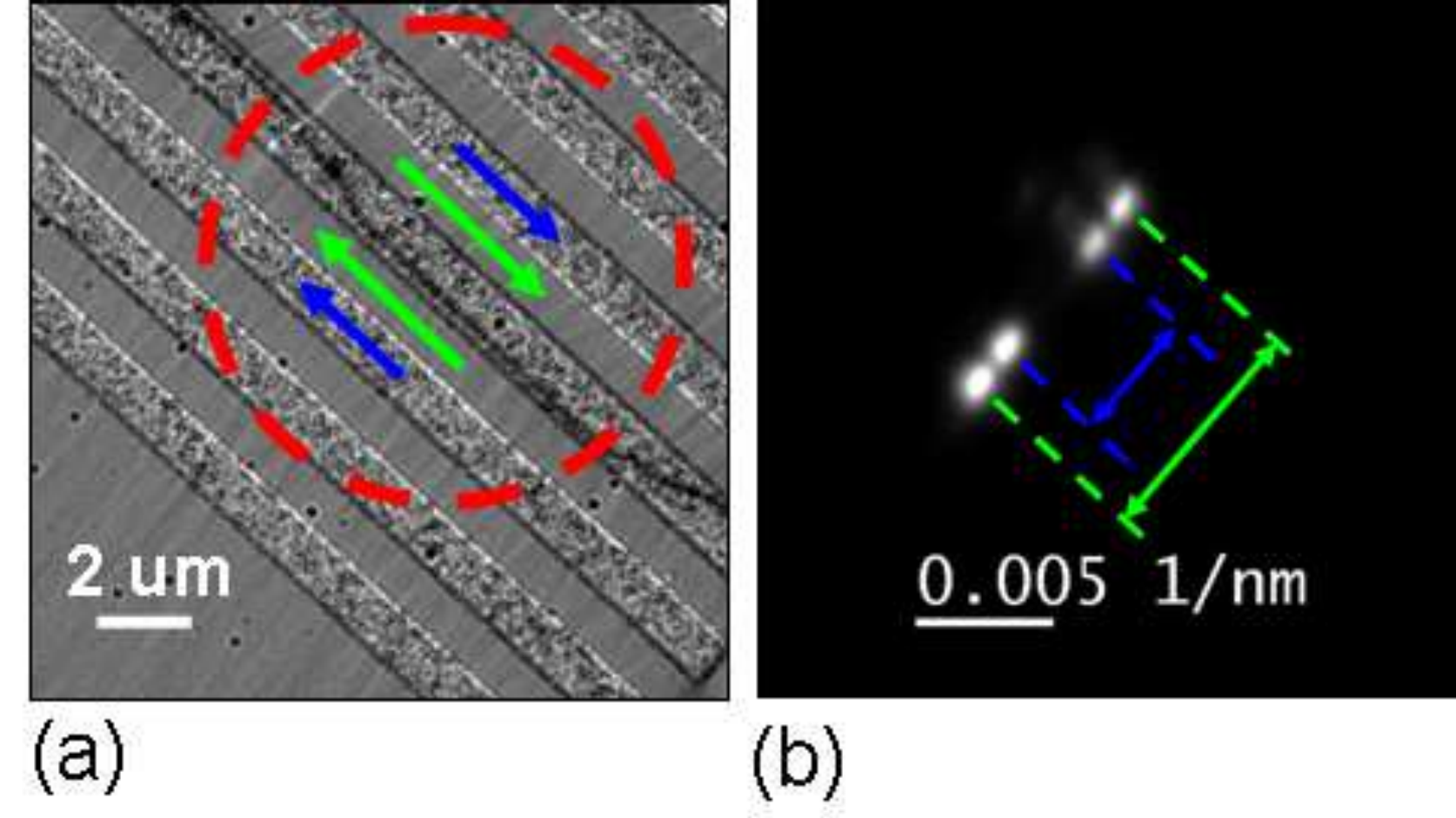}
\caption{(a) Fresnel image showing the oppositely magnetised magnetic region of the unirradiated and irradiated stripes. Green and blue arrows denote magnetization in unirradiated and irradiated striped respectively. An $180^{\circ}$ domain wall (black line in image) is clearly visible along the stripe pattern between the regions of opposite magnetization. (b) Electron diffracted spots were obtained by illuminating the area, marked by red circle in image (a), from low angle diffraction experiments using Lorentz TEM. The two outer spots (indicated by the green marking) arise from the unirradiated regions whereas two inner spots (indicated by blue)  are from the irradiated regions.} \label{fig2}
\end{figure}

\section{Results and discussions} \label{III}
\subsection{Structural and magnetic characterization} \label{I}
As an illustration of the changes in both film morphology and magnetic properties due to local ion irradiation we present here an exemplary detailed study for the case of FIB irradiation. The plan-view bright field (BF) TEM image in \textit{figure \ref{fig1}(b)} shows both irradiated and non-irradiated stripes with a width of $1000 \ nm$. The non-irradiated and irradiated regions can easily be distinguished as the irradiation has caused considerable increase in the grain size from  $5-10 \ nm$ to $30-40 \ nm$. An increase of grain size after $Ga^+$ focused ion beam irradiation of $NiFe$ films has also been reported earlier \cite{ref11,ref12,ref13,ref14}. A cross-sectional TEM BF image of a stripe pattern of $200 \ nm$ stripe width, located on a frame of a silicon-nitride membrane, is shown in \textit{figure \ref{fig1}(c)}. The irradiated layer stack can be recognized by the reduced thickness due to irradiation induced sputtering \textit{figure \ref{fig1}(c)}. The actual thickness deposited and of the irradiated $Ni_{81}Fe_{19}$ layer has been determined to be $16.9\pm 0.5$ $nm$ and $14\pm 0.5$ $nm$, respectively. TRIDYN \cite{ref15} simulations of the irradiation process of the layer system $Ni_{81}Fe_{19}(16.9nm)/Si_3N_4(25nm)/Si$ suggest an inhomogeneous Gallium profile with a maximum concentration of $5.1 \%$ at a penetration depth of $8.7 \ nm$ and a surface recession of $2 \ nm$. The latter is in reasonable agreement with the measured changes in film thickness. The magnetization configuration during external field reversal inside the stripe patterns was imaged by means of Lorentz microscopy \cite{ref18, ref19} in the Fresnel and differential phase contrast (DPC) mode. The DPC mode is practised on a scanning transmission electron microscope (STEM). Initially low angle electron diffraction \cite{ref19, ref20} experiments were carried out in the TEM to measure the difference of the product of saturation induction B$_S$ and film thickness t, B$_S$t, between the unirradiated and irradiated areas of the film. Using the Fresnel mode of Lorentz TEM, a $180^{\circ}$ domain wall was formed along the stripe pattern during reversal so that oppositely magnetized domains in both regions were present as shown in figure 2(a), where the wall is visible as a black line. Fine scale structure due to magnetization ripple can been seen, this runs perpendicular to the direction of magnetization allowing arrows to be assigned either side of the wall. Therefore, pairs of diffraction spots were formed by illuminating the area, marked by red circle, of the anti-parallel alignment of magnetization, \textit{figure \ref{fig2}(a)}. In \textit{figure \ref{fig2}(b)} four diffraction spots are observed for the stripe pattern, two from oppositely magnetized non-irradiated regions (green marked) and the other two from the oppositely magnetized regions with reduced saturation magnetization value and film thickness (blue marked). The ratio of the values of $B_S t$ (and hence magnetic moment) between the irradiated and unirradiated stripes is $60$\%. For a decrease in film thickness of  $2.9 \ nm$  as obtained from cross-sectional TEM measurements, the magnetic induction $B_S$ is calculated to be $72$\% of the value in the unirradiated, i.e. as deposited film.  

\subsection{Stripe width 1000 $nm$} \label{II}

\begin{figure}[!t]
\centering
\includegraphics[width=6.5 cm]{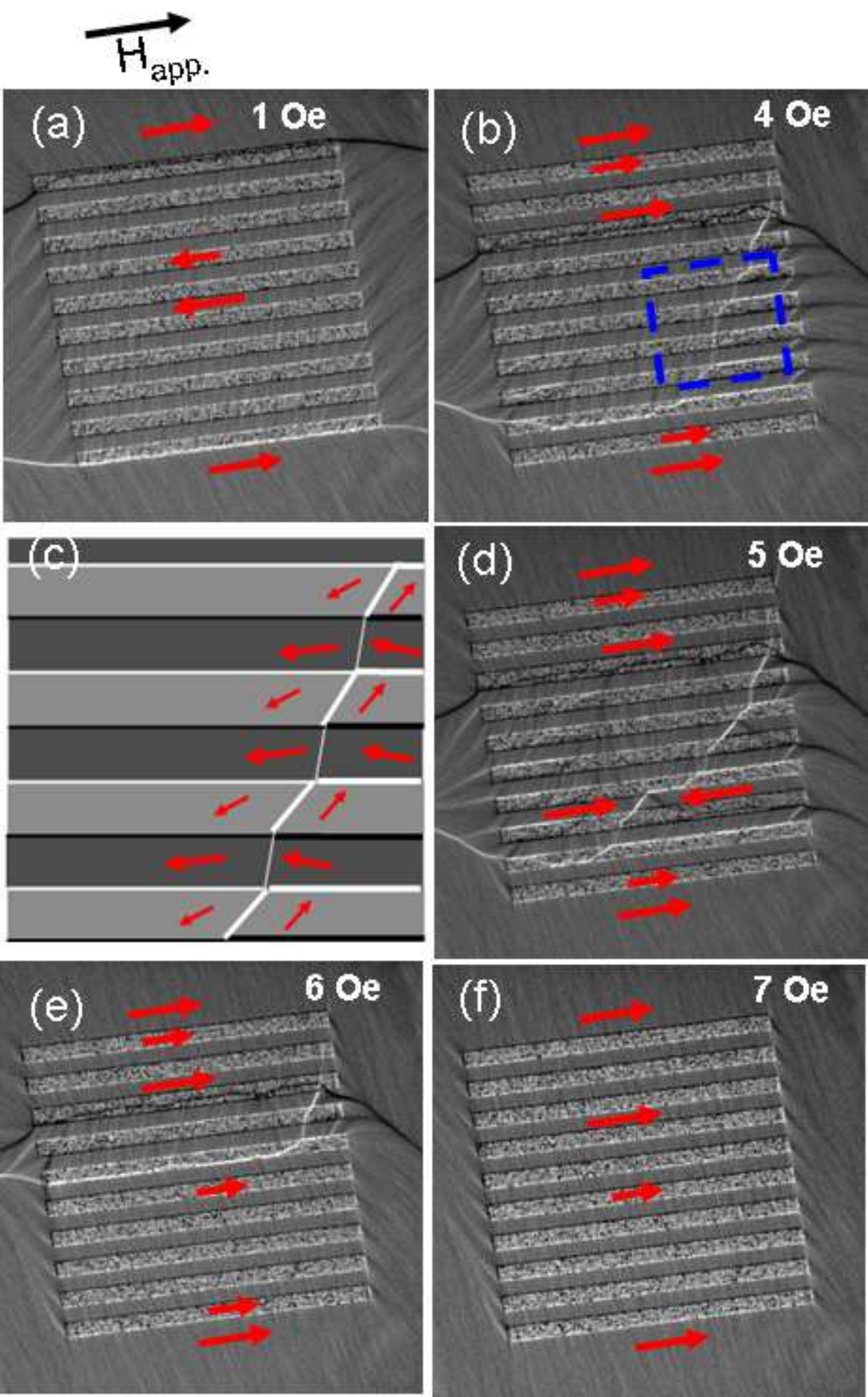}
\caption{Fresnel image sequence (a,b,d-f) of the magnetization reversal inside the magnetization modulated film. (c) is a schematic of blue marked region of Fresnel image (b). The irradiated stripes are shown in lighter grey. The nominal stripe width is $1000 \ nm$.} \label{fig3}
\end{figure}

The magnetization reversal inside the modulated film was first studied for a stripe width of $1000 \ nm$ by Fresnel imaging and is shown in figure 3. The sample was initially saturated along the long axis of the stripe with a negative external field which was then reduced to zero. Subsequently the field was then increased in the positive direction. In \textit{figure \ref{fig3}}, with the field at 1 Oe, it can be seen that the continuous film region has reversed whilst the magnetization in the patterned region remains in the negative direction. This is apparent as domain walls are clearly visible at either side of the long edges of the patterned region. From the magnetization ripple visible (not apparent at displayed size) inside the as-prepared soft magnetic $Ni_{81}Fe_{19}$ film, the magnetization orientation is deduced (\textit{large red arrows}). The magnetization direction inside the irradiated material (\textit{small red arrows}) cannot be deduced directly, as no sufficient magnetization ripple contrast is present there. Therefore the magnetization direction is calculated based on magnetic-flux density continuation arguments, assuming a continuous magnetic induction perpendicular across the stripes. Due to different film thickness and different saturation magnetization values an alternating dark and bright Fresnel contrast is apparent at the interfaces between the non-irradiated and irradiated stripes (this is very clearly visible in \textit{figure \ref{fig2}(a)}). The  domain walls separating the continuous film and the stripe pattern have much higher contrast than the lines between the unirraiated and irradiated regions as the walls have a $180^{\circ}$ change in magnetization direction whereas between the stripe region the magnetization changes between different strengths of parallel magnetization. 

On increasing the strength of the applied field to 4 $Oe$, figure \textit{\ref{fig3}(b)}, the irradiated stripes start to reverse their magnetization. Spike-like domains nucleate at the stripe's ends and the related walls propagate along the stripes. Similar to extended hybrid stripe patterns \cite{ref2}, a collective reversal of magnetization of the irradiated stripes was observed as shown in the blue square marked region of \textit{figure \ref{fig3}(b)} and it's schematic \textit{figure \ref{fig3}(c)}. These domain walls separate areas of nearly opposite magnetization orientations inside the irradiated stripes, being high angle walls. But associated walls are also present in the non-irradiated stripes. There they separate areas with magnetization orientations slightly tilted in respect to each other, being low angle walls. Therefore, areas of neighboring stripes are seen to exist with opposite longitudinal magnetization components, separated by high angle domain walls at the stripe interfaces (\textit{figure \ref{fig3}(b)}). This is also indicated in the schematic \textit{figure \ref{fig3}(c)} where the stripe interfaces are shown to have strong bright and dark contrast in this region, consistent with the contrast in the image in \textit{figure \ref{fig3}(b)}. Furthermore in \textit{figure \ref{fig3}(b)} the high angle domain walls first seen in \textit{figure \ref{fig3}(a)} extending into the surrounding film and formerly located at the outer irradiated stripes, are observed to have moved already to the inside of the stripe pattern. Due to this movement the magnetization has reversed completely in all the stripes traversed by these walls. It should be noted that these walls are only ever seen in the irradiated stripes and not in the non-irradiated stripes.  Obviously it is energetically more favourable to locate these walls in the irradiated material due to smaller values of both saturation magnetization and film thickness. Upon further increase of the external field value the magnetization inside the non-irradiated stripes starts reversing before the irradiated stripes have completed their reversal (\textit{figure \ref{fig3}(d)}). Domain walls propagating along the non-irradiated stripes are pinned and transverse domain walls are formed only inside the non-irradiated stripes (\textit{figure \ref{fig3}(d)}). The transverse walls are similar in form to walls observed in patterned nanowires \cite{ref14}. The reversal of the hybrid structure is influenced by the high angle domain walls in the surrounding film which were initially pinned at the outer structure regions. These domain walls move when the external field exceeds the appropriate depinning fields (\textit{figure \ref{fig3}(e)}) and the hybrid structure finally completes the magnetization reversal (\textit{figure \ref{fig3}(f)}). 

Looking at the reversal process of the stripes, it can be determined whether the unirradiated or irradiated stripes with high or low saturation magnetization value, respectively start the magnetization reversal. The increased coercivity in the irradiated stripes with low saturation magnetization value support a switching of the unirradiated stripes with high saturation magnetization value. Notably, changes of the magnetization reversal along the initial easy axis were measured for $70 \ \mu m$ x $70 \ \mu m$ squares of a $Ni_{81}Fe_{19}(20nm)/Si_3N_4$ reference sample irradiated with focused ion beam with the same parameters. The coercivity H$_c$ increased to at least 6 $Oe$ due to an increase in pinning of the domain walls. However, by considering the difference in energy between the two regions, it would be expected that the energy associated with the irradiated region is lower due to lower moment and possibly exchange. Therefore, a switching of these irradiated stripes first is expected. Obviously this seems to be the dominant part and the fact that the DWs reside in the irradiated stripe is an indication of this. To support this argument, micromagnetic simulations were carried out and results will be presented in \textit{section D}.

\begin{figure}[!hh]
\centering
\includegraphics[width=7cm]{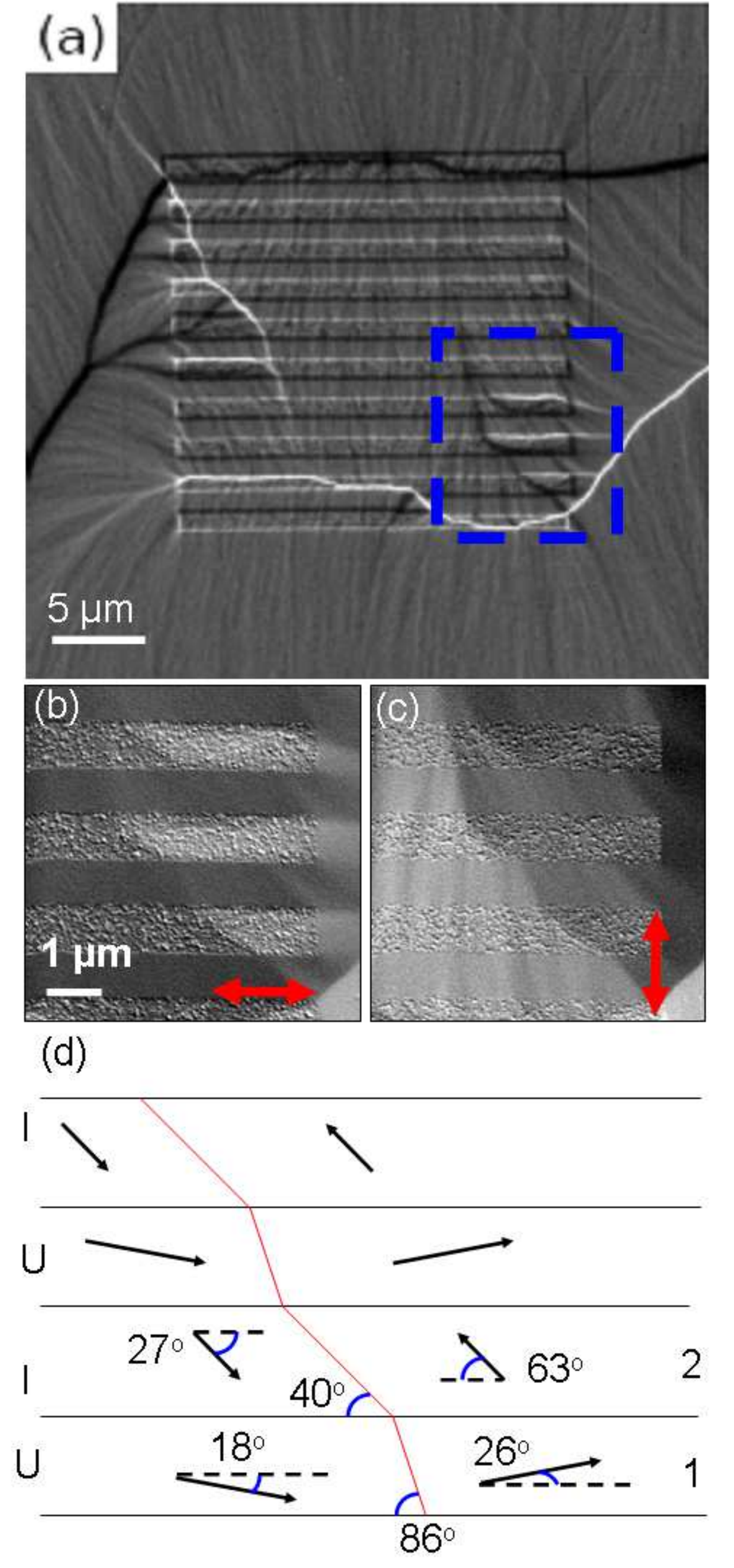}
\caption{High resolution differential phase contrast imaging of a magnetization state during field reversal.  The two sensitivities are shown by double headed arrows. The stripe width is $1 \ \mu m$. The wall contrast of the Fresnel image (a) can be directly compared with both magnetic induction components (b and c) mapped by DPC imaging. Induction components in (b) and (c) are shown schematically in (d).} \label{fig4}
\end{figure}

Details of the magnetization orientations of the formerly described magnetization pattern of \textit{figure \ref{fig3} (b)} are now discussed. The magnetization reversal cycle was repeated in order to produce a similar domain geometry as shown in Fresnel image \textit{figure \ref{fig4} (a)}. By switching from Fresnel to DPC imaging and then imaging the area indicated in \textit{figure \ref{fig4} (a)}, the magnetic induction components parallel (\textit{figure \ref{fig4} (b)}) and perpendicular (\textit{figure \ref{fig4} (c)}) to the stripe axis are mapped. In the lower right edge of the panel in figure 4(a) (blue rectangle) again a reversal in part of the irradiated material can be observed to have started. The mapped magnetic induction component perpendicular to the stripe axis (\textit{figure \ref{fig4} (c)}) shows a transverse magnetization component in both irradiated and non-irradiated stripes. Despite this transverse magnetization component, periodic anti-parallel magnetization components along the stripe axis are present (\textit{figure \ref{fig4} (b)}). Magnetization components in \textit{figure \ref{fig4}(b)} and \textit{(c)} are shown schematically in \textit{figure \ref{fig4}(d)} along with the rotation angles measured from these DPC images. The calculation was based on consistency between the signal levels on the two components, the wall angles and the assumption of no net charge on the walls and interfaces. As shown in Fresnel image (\textit{figure \ref{fig3}(d)}) where the reversal started in the irradiated stripes, what happens in the application of a reverse field is that there is a rotation of the magnetisation in the stripes. At low fields this is fairly small in the non-irradiated stripes. If there is to be no charge at the interface then there is a much larger rotation in the irradiated stripes and we get a larger angle DW there. This is in fact what we see from the schematic shown in \textit{figure \ref{fig4}(d)}. If we consider the unirradiated and irradiated stripes 1 and 2, respectively in the schematic \textit{figure \ref{fig4}(d)}, it is clear that this state is consistent with no net charge being present at the interfaces and walls. Therefore, a larger rotation of magnetization in the irradiated stripe is expected by forming high angle DWs.

In our experiments we were not able to observe a complete anti-parallel magnetization alignment in neighboring stripes. In such a case the tails of two neighboring N\'eel walls (located at the interfaces of the stripes) must fit into the dimension of one stripe's width. Obviously the dimension of the tails are much larger. So the high angle domain walls between neighboring stripes are densely packed, but the overlapping tails of the N\'eel walls do not allow for a complete anti-parallel alignment.

\subsection{Stripe width $200 \ nm$ } \label{IIIB}

\begin{figure}[!hh]
\centering
\includegraphics[width=7cm]{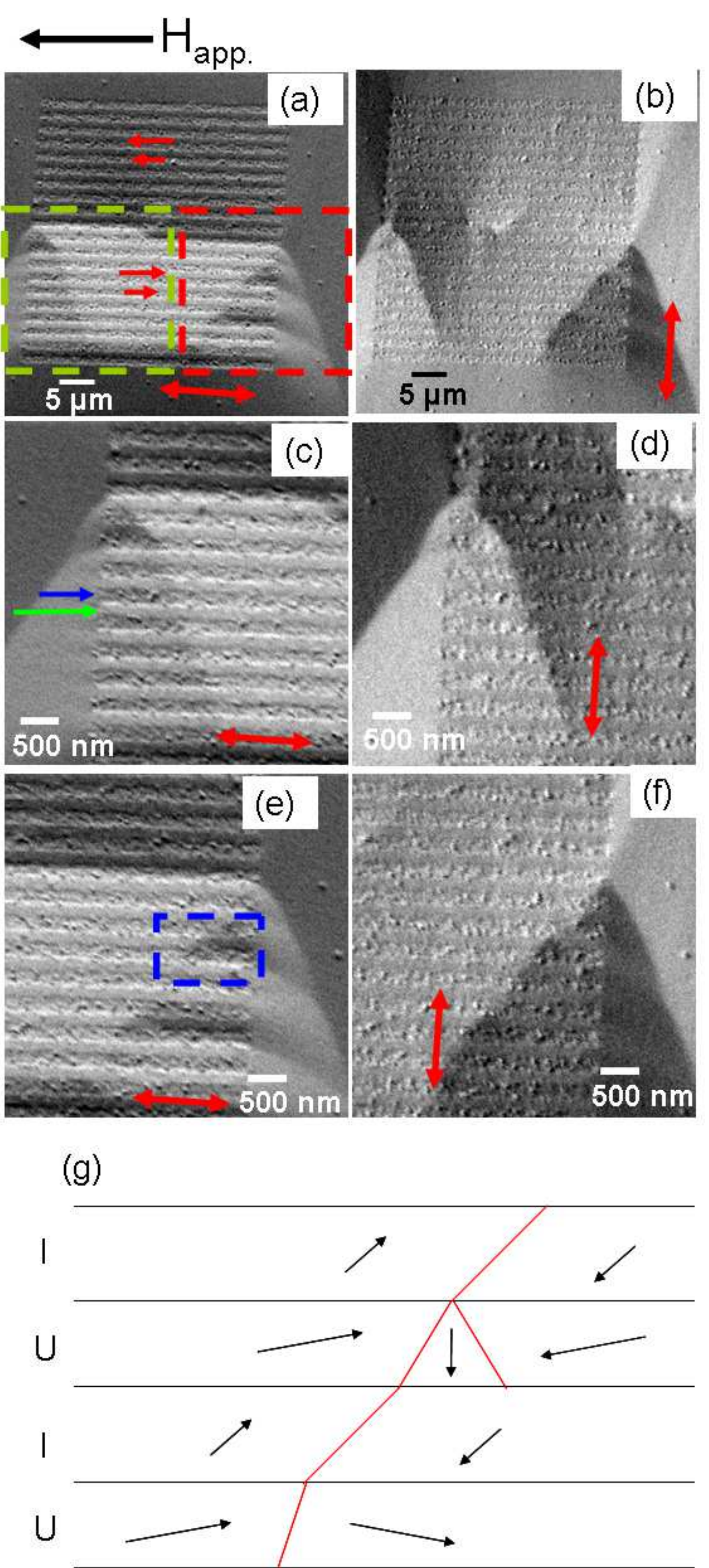}
\caption{DPC images of stripe array with stripe width of $200 \ nm$. The two sensitivities are shown by double headed arrows. Images (a,b) show the magnetic induction distribution for the whole hybrid structure. Images c and d (e and f) are showing the magnification of the green and red rectangular regions, respectively of image (a). The two arrows in image (c) denote the unirradiated (green) and irradiated (blue) stripes. (g) is the schematic of the marked region in DPC image (e).} \label{fig5}
\end{figure}

The effect of increasing the interaction of the densely packed high angle domain walls between the stripes on the magnetization reversal of the hybrid structure is studied by scaling the patterning size down to a stripe width of $200 \ nm$. In \textit{figure \ref{fig5}} differential phase contrast images for a representative domain state during external magnetic field reversal of a $200 \ nm$ wire width are shown. The mapped magnetic induction component is parallel or longitudinal (perpendicular or transverse) to the stripes are shown by double headed arrows. 

The pair of images \textit{figures \ref{fig5}(a,b)} show the magnetic induction distribution for the whole hybrid structure. The upper situated stripes as indicated by the red arrows, \textit{figure \ref{fig5}} (a) can be seen to have already completed reversal and these stripes are magnetized along the external field direction. In pair of images \textit{figures \ref{fig5}(c,d)} a magnified region of the left lower part (green rectangle) of the structure is shown. Here, comparing the longitudinal and transverse magnetization component, it is evident that the low angle wall separates areas with slightly different magnetization orientations. Not only the magnetization inside the stripes with low saturation magnetization value is rotating away from the stripe axis, but also the magnetization in the stripes with high value follow this rotation for the reason of flux continuity. Also from the angle of low angle wall in respect to the stripe axis it could be deduced that the magnetization at the stripes ends have a transverse component. Apparently no discrete pattern with anti-parallel alignments of the magnetization of neighboring stripes is found which was observed for wider stripes in \textit{figures \ref{fig3}} and \textit{figure \ref{fig4}}. In pair of images \textit{figures \ref{fig5}(e,f)} a magnification into the right lower part (red rectangle) of the structure is shown. Compared to the left structure side, the domain wall is a high angle wall for two irradiated stripes inside the blue rectangular region of \textit{figure \ref{fig5}(e)}. The magnetization of this blue rectangle region is shown in the schematic diagram, \textit{figure \ref{fig5}(g)}. Inside the two irradiated stripes the magnetization has opposite longitudinal components on both sides of the wall. Therefore high angle domain walls are formed at the interfaces of the reversed irradiated stripe and the neighboring unirradiated stripe. In the unirradiated stripes we see that both a head to head wall (upper stripe) is supported and a low angle wall (lower stripe) can be supported. But periodic neighboring high angle domain walls between adjacent stripes which were observed for wider stripes (1000 $nm$) are not found in these narrower stripes. Instead of this the stripes tend to reverse together. Therefore the patterning size provides an effective magnetic medium as was reported for  $2 \ \mu m$ width stripe pattern embedded in a weakly antiferromagnetically exchange coupled 
$Co_{90}Fe_{10}$/$Ru$/$Co_{90}Fe_{10}$ trilayer system \cite{ref64}. Notably in Ref. [12], the magnetic reversals behavior of micrometer size stripe patterns achieved by ion irradiation in a weakly antiferromagnetically exchange coupled $Co_{90}Fe_{10}$/$Ru$/$Co_{90}Fe_{10}$ system was explained by modification of magnetic material properties, e.g. anisotropy and saturation magnetization. Although the stripe width was much larger i.e. $2 \ \mu m$ compared to that of 200 $nm$ fabricated in our system, however, that observation indicates interactions of the internal magnetization of embedded stripes with the adjacent ones in the nonirradiated antiferromagnetically coupled trilayers. In our investigation, the formation of densely packed high angle N\'eel walls between the narrower stripes is therefore suppressed for this structure size.

\subsection{Micromagnetic simulation on stripe patterns}

The two-step reversal of the irradiated stripes of width 1000 $nm$ is attributed to the lower energy of the domain walls in this region. To support this argument, micromagnetic simulations using the OOMMF package \cite{ref53} were carried out for an assembly of two neighboring stripes with different values of saturation magnetization $M_S$ and exchange stiffness constant $A$. The dimensions of each stripe and the cell size of the simulation are chosen to be $4 \ \mu m \times 200 \ nm \times 20 \ nm $ and $5 \times 5 \times 5 \ nm^3 $, respectively. Inside the simulation the magnetic material parameters are used as follows: saturation magnetization $M_{S,1}$ = $8.6 \times 10^{6} \ A/m$, exchange stiffness constant $A_{11} = 1.3 \times 10^{-11} \ J/m$ and anisotropy constant $K$ =0. A damping coefficient $\alpha$ of $0.5$, though not typical for $Ni_{81}Fe_{19}$ is used in order to speed up the relaxiation of the micromagnetic configurations. The saturation magnetization value of the neighboring stripe with reduced values is $M_{S,2}$ at $60\%$ compared to the initial value of $M_{S,1}$. The exchange stiffness constant $A_{22}$ of the stripe with reduced saturation magnetization value and the exchange coupling stiffness constant between the stripes $A_{12}$ are estimated with the relation \textit{\ref{exchange1}} and \textit{\ref{exchange2}} \cite{ref52}.

\begin{equation}
A_{22} = A_{11}(\frac{M_{s,irradiated}}{M_{s,unirradiated}})^2= 4.68\times 10^{-12} \ J/m
\label{exchange1}
\end{equation}

\begin{equation}
A_{12} = A_{21} = 0.5 (A_{11}+A_{22})= 8.84\times 10^{-12}J/m
\label{exchange2}
\end{equation}

\begin{figure}[!hh]
\centering
\includegraphics[width=8cm]{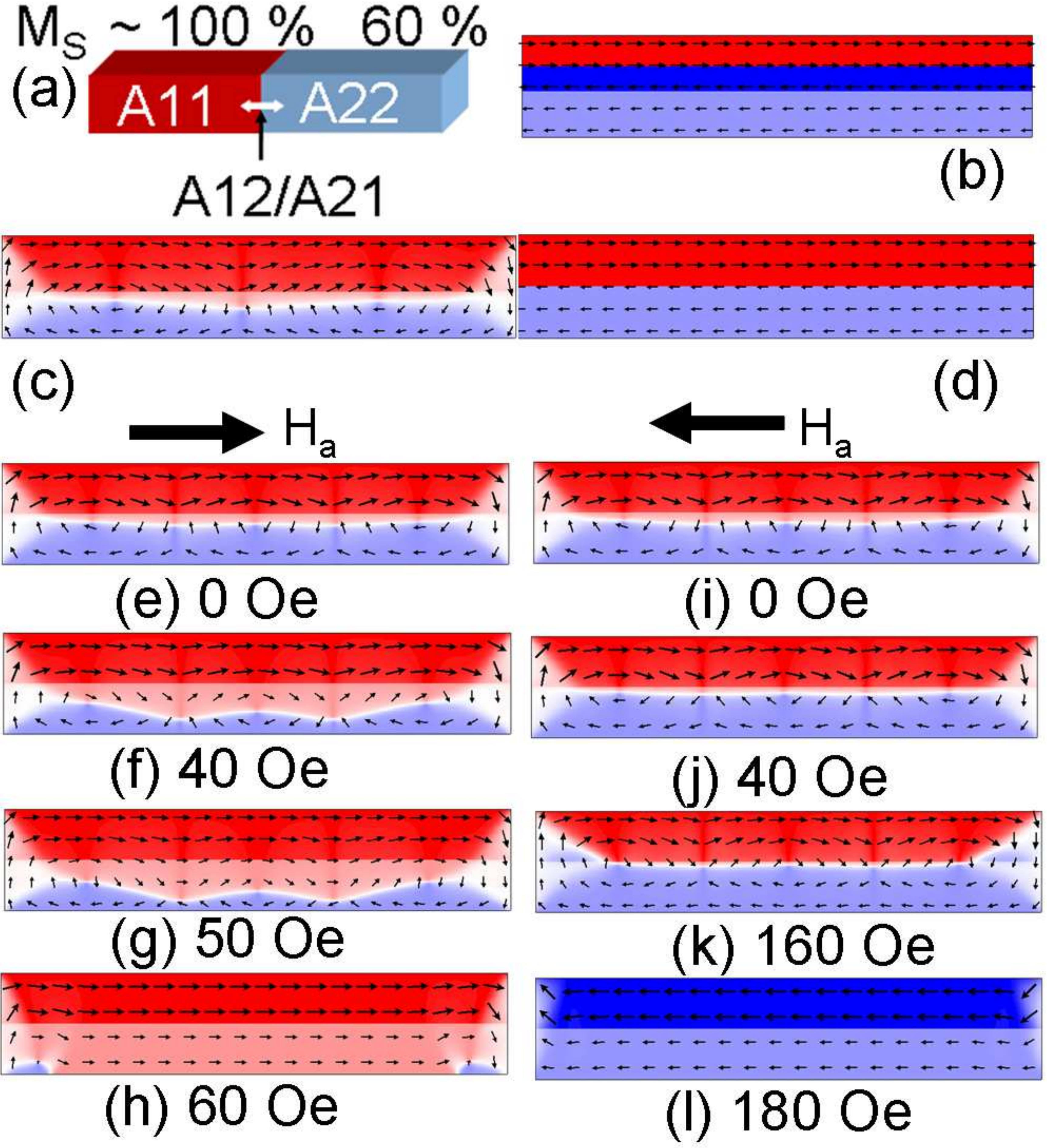}
\caption{Schematic cross-section (a) of a stripe assembly simulated in OOMMF with different saturation magnetization values $M_S$, exchange stiffness constants $A_{11}$, $A_{22}$ and coupling constant $A_{12}$. After initalizing the magnetization states (b and d) the magnetization relaxes into an flux closure pattern (c and e,i). The movement of the formed high angle domain wall near the interface of the two stripes for two different external field directions $H_a$ is shown in (e-h and i-l). The dimensions of each stripe and the cell size of the simulation are chosen to be $4 \ \mu m \times 200 \ nm \times 20 \ nm $ and $5 \times 5 \times 5 \ nm^3 $, respectively.} \label{fig7}
\end{figure}

A schematic cross-section of the two neighboring stripes is illustrated in \textit{figure \ref{fig7}(a)}. First an inital magnetization configuration (\textit{figure \ref{fig7}(b)}) with a $180^{\circ}$ domain wall in the stripe with saturation magnetization $M_{S,1}$ was set and relaxed at zero external field to the domain state shown in \textit{figure \ref{fig7}(c)}.

The initially configured $180^{\circ}$ domain wall is not stable and a flux closure pattern is formed on relaxation, hosting a high angle domain wall near the interface of both stripes. This domain wall is located in the stripe with reduced saturation magnetization value $M_{S,2}$ due to the lower wall energy and not at the interface itself. For the rest of the simulations an initial magnetization state with a $180^{\circ}$ domain wall at the interface of both stripes (\textit{figure \ref{fig7}(d)}) is used, where the relaxed state (\textit{figure \ref{fig7}(e,i)}) is almost equal to the previous state (\textit{figure \ref{fig7}(c)}). The displacement of the high angle domain wall upon the application of a magnetic field along the stripe axis for two opposite field directions is studied, see \textit{figure \ref{fig7}(e-h)} and \textit{\ref{fig7}(i-l)}. It is apparent that the domain wall is free to move in the stripe (with changed magnetic parameters), reversing finally the magnetization in this stripe in the applied field direction (\textit{figure \ref{fig7}(e-h)}). Applying the field anti-parallel to the magnetization of the lower stripe (\textit{figure \ref{fig7}(j-l)}) with reduced saturation magnetization $M_{S,2}$ in the relaxed state, the domain wall gets pinned close to the interface between the stripes (\textit{figure \ref{fig7}(j-k)}). In this case, the reversal of the magnetization in the stripe with saturation magnetization value $M_{S,1}$ is completed by rotation of the moments inside the stripe instead of moving the high angle domain wall across the energy barrier. This different field direction dependent behaviour confirms the impact of the reduction in wall energy due to the reduction of the saturation magnetization value on the reversal mechanisms. This ultimately supports the argument that irradiated stripe start reversal process during the experiments.

\section{Conclusions} \label{II}
Saturation magnetization modulated stripes embedded in a $Ni_{81}Fe_{19}$ matrix were investigated regarding the occurrence of anti-parallel magnetization components for adjacent stripes. Starting from a stripe width of 1000 $nm$ a pronounced two step reversal with nearly anti-parallel orientated magnetization in neighboring stripes is observed. It is evident that there is no net charge present at the interfaces and walls. Therefore, a larger rotation of magnetization in the irradiated stripe is observed by forming high angle DWs. However, the spacing between the neighboring nearly $180^{\circ}$ domain walls still allows for discrete stripe domain patterns. By scaling the patterning size down to a stripe width of 200 $nm$, we observe a transition in the characteristic magnetization patterns during external field reversal. Due to strong domain wall tail interactions in adjacent stripes the formation of neighboring high angle DWs at the interfaces of these stripes is suppressed. Therefore a tendency of collective reversal of the stripe pattern is observed. We have also shown the power of Lorentz microscopy to image structures well under the micron length-scale allowing the observation of dimensions not covered by previous investigations \cite{ref2, ref3,ref64}. Therefore the potential for future applications in patterned nanoscale structures of these materials for sensors and spintronic devices may be demonstrated using such a technique.



\begin{thebibliography}{99}

\bibitem{ref81} M. Diegel, S. Glathe, R. Mattheis, M. Scherzinger, and E. Halder, IEEE Trans. Magn. 45, 3792 (2009).

\bibitem{ref82} B. C. Stipe, T. C. Strand, C. C. Poon, H. Balamane, T. D. Boone, J. A. Katine, J.-L. Li, V. Rawat, H. Nemoto, A. Hirotsune, O. Hellwig, R. Ruiz, E. Dobisz, D. S. Kercher, N. Robertson, T. R. Albrecht, and B. D. Terris, Nat. Photon. 4, 484 (2010).

\bibitem{ref1} L. Folks, R. E. Fontana, B. A. Gurney, J. R. Childress,
S. Maat, J. A. Katine, J. E. E. Baglin, and A. J. Kellock,
Journal of Physics D: Applied Physics, 36(21), 2601{2604}
(2003). 
\bibitem{ref51} D. McGrouther and J. N. Chapman, App. Phys. Lett., 87, 022507, (2005).
\bibitem{ref2} J. McCord, L. Schultz, and J. Fassbender, Advanced Ma-
terials, 20(11), 2090 (2008). 
\bibitem{ref3} J. Fassbender, T. Strache, M. O. Liedke, D. Marko,
S. Wintz, K. Lenz, A. Keller, S. Facsko, I. Monch,
and J. McCord, New Journal of Physics, 11(12), 125002
(2009). 
\bibitem{ref61} I. Barsukov, F. M. R\"omer, R. Meckenstock, K. Lenz, J. Lindner,
S. Hemken to Krax, A. Banholzer, M. K\"orner, J. Grebing, 
J. Fassbender, and M. Farle, Physical Review B 84, 140410(R) (2011).
\bibitem{ref4} S. I. Woods, S. Ingvarsson, J. R. Kirtley, H. F. Hamann,
and R. H. Koch, Applied Physics Letters, 81(7), 1267{1269}
(2002). 
\bibitem{ref5} J. McCord, T. Gemming, L. Schultz, J. Fassbender, M. O.
Liedke, M. Frommberger, and E. Quandt, Applied Physics
Letters, 86(16), 162502 (2005). 
\bibitem{ref6} J. McCord, I. Moench, J. Fassbender, A. Gerber, and
E. Quandt, Journal of Physics D: Applied Physics, 42(5),
055006 (2009). 
\bibitem{ref62} S. C. Shin, S. K. Han, J. Hong, and S. Kang, Appl. Phys. Express
4, 116501 (2011).
\bibitem{ref64} M. Langer, A. Neudert, J. I. M\"onch, R. Mattheis, K. Lenz, J. Fassbender, and J. McCord, Physical Review B 89, 064411 (2014).
\bibitem{ref63} C. Hamann, R. Mattheis, I. M\"onch, J. Fassbender, L. Schultz and J. McCord, New Journal of Physics, 16, 023010 (2014).
\bibitem{ref71} M. A. Basith, S. McVitie, D. McGrouther and J.N. Chapman, Applied Physics Letters, 100, 232402, (2012).
\bibitem{ref83} M. J. Benitez, M. A. Basith, R. J. Lamb, D. McGrouther, S. McFadzean, D. A. MacLaren, A. Hrabec, C. H. Marrows, and S. McVitie, Physical Review Applied, 3, 034008 (2015).
\bibitem{ref7}	A. Hubert and R. Schaefer, Magnetic Domains, Springer,
1998 [Chapter 3.6.4, pages 223-241].

\bibitem{ref9} P.-O. Jubert, R. Allenspach, and A. Bischof, Physical Re-
view B, 69(22), 220410 (2004).
\bibitem{ref10} M. Ruehrig, B. Khamsehpour, K. Kirk, J. Chapman,
P. Aitchison, S. McVitie, and C. Wilkinson, IEEE Trans-
actions on Magnetics, 32(5), 4452{4457} (1996). 
\bibitem{ref11} C.-M. Park and J. Bain, IEEE Transactions on Magnet-
ics, 38(5), 2237{2239} (2002). 
\bibitem{ref12} C.-M. Park and J. A. Bain, Journal of Applied Physics,
91(10), 6830{6832} (2002).
\bibitem{ref13} D. Ozkaya L, R. M. Langford, W. L. Chan, and A. K.
Petford-Long, Journal of Applied Physics, 91(12), 9937{9942} (2002). 
\bibitem{ref14} M. A. Basith, S. McVitie, D. McGrouther, J. N. Chap-
man, and J. M. R. Weaver, Journal of Applied Physics,
110, 083904 (2011).
\bibitem{ref15} W. Moeller, W. Eckstein, and J. P. Biersack, Computer
Physics Communications, 51(3), 355{368} (1988).
\bibitem{ref16} M. Goto, H. Tange, and T. Kamimori, Journal of Mag-
netism and Magnetic Materials, 62(2-3), 251{255} (1986).
\bibitem{ref17} D. Marko, T. Strache, K. Lenz, J. Fassbender, and
R. Kaltofen, Applied Physics Letters, 96(2), 022503 (2010).
\bibitem{ref18} J. N. Chapman, A. B. Johnston, L. J. Heyderman,
S. McVitie, W. A. P. Nicholson, and B. Bormans, IEEE
Transactions on Magnetics, 30(6), 4479{4484} (1994).
\bibitem{ref19} J. N. Chapman, Journal of Physics D: Applied Physics,
17(4), 623 (1984).
\bibitem{ref20} Y. Togawa, T. Koyama, K. Takayanagi, S. Mori, A. Kousaka, J. Akimitsu, S. Nishihara, K. Inoue, A. S. Ovchinnikov, and J. I. Kishine, Phys. Rev. Lett. 108, 107202 (2012).
\bibitem{ref53} M. J. Donahue and D. G. Porter Report no. nistir 6376
Technical report, National Institute of Standards and
Technology, Gaithersburg, MD, (1999).
\bibitem{ref52} R. Schaefer, Journal of Magnetism and Magnetic Materials, 215-216, 652 (2000).



\end{thebibliography}
\end{document}